\documentclass[aps,prd,nofootinbib,superscriptaddress,preprint,eqsecnum,showkeys,showpacs,preprintnumbers]{revtex4}

\usepackage{color}
\usepackage{slashed}
\usepackage{graphics}
\usepackage{graphicx}
\usepackage{dcolumn}
\usepackage{subfigure}
\usepackage{mathrsfs}
\usepackage{bm}
\usepackage{amsmath,amssymb,epsfig}
\usepackage{float}

\newcommand{\pa}{\partial}

\begin{document}

\title{New self-consistent effective one-body theory for spinless  binaries based on the post-Minkowskian approximation}

\author{Jiliang {Jing}\footnote{ jljing@hunnu.edu.cn}}
 \affiliation{Department of Physics, Key Laboratory of Low Dimensional Quantum Structures and Quantum Control of Ministry of Education, and Synergetic Innovation
Center for Quantum Effects and Applications, Hunan Normal
University, Changsha, Hunan 410081, P. R. China}
\affiliation{Center for Gravitation and Cosmology, College of Physical Science and Technology, Yangzhou University, Yangzhou 225009, P. R. China}

\author{Sheng Long}
 \affiliation{Department
of Physics, Key Laboratory of Low Dimensional Quantum Structures and
Quantum Control of Ministry of Education, and Synergetic Innovation
Center for Quantum Effects and Applications, Hunan Normal
University, Changsha, Hunan 410081, P. R. China}

\author{Weike Deng}
\affiliation{Department
of Physics, Key Laboratory of Low Dimensional Quantum Structures and
Quantum Control of Ministry of Education, and Synergetic Innovation
Center for Quantum Effects and Applications, Hunan Normal
University, Changsha, Hunan 410081, P. R. China}

\author{Mengjie Wang\footnote{mjwang@hunnu.edu.cn}}
\affiliation{Department
of Physics, Key Laboratory of Low Dimensional Quantum Structures and
Quantum Control of Ministry of Education, and Synergetic Innovation
Center for Quantum Effects and Applications, Hunan Normal
University, Changsha, Hunan 410081, P. R. China}

\author{Jieci Wang\footnote{ jcwang@hunnu.edu.cn}} 
\affiliation{Department
of Physics, Key Laboratory of Low Dimensional Quantum Structures and
Quantum Control of Ministry of Education, and Synergetic Innovation
Center for Quantum Effects and Applications, Hunan Normal
University, Changsha, Hunan 410081, P. R. China}


\begin{abstract}
The effective one-body theories, introduced by Buonanno and Damour, are novel approaches to constructing a gravitational waveform template. By taking a gauge in which $\psi_{1}^{B}$ and $\psi_{3}^{B}$ vanish, we find a decoupled equation with separable variables for $\psi^{B}_{4}$ for gravitational perturbation in the effective metric  obtained in the post-Minkowskian  approximation. Furthermore, we set up a new self-consistent effective one-body theory for spinless binaries, which can be applicable to any post-Minkowskian orders. This theory not only releases the assumption that $v/c$ should be a small quantity but also resolves the contradiction that the Hamiltonian, radiation-reaction force, and waveform are constructed from different physical models in the effective one-body theory with the post-Newtonian approximation.
Compared with our previous theory (Science China, 65, 260411, (2022)), the computational effort for the radiation-reaction force and waveform in this new theory will be tremendously reduced.

\end{abstract}


\maketitle

 \section{Introduction}
The investigation on gravitational waves (GWs) has attracted much attention since 1918 \cite{einstein18,BonVanMet62,Sac62, deAdeBoc00,KomSmiDun11,PerAldDel98,Jing2021,Jing2019}, and
many direct detections of GWs have been recently announced by LIGO, Virgo, and KAGRA ~\cite{Abbott2016,Abbott2016(2),Abbott2017,Abbott2017(2),Abbott2017(3),Abbott2019,Abbott20211,Abbott20212,Abbott20213,Nitz2021}. The success of direct detections is based on the development of technology and theoretical research. In theoretical studies, a gravitational waveform template (GWT) plays a central role. The basis of the GWT is to study the late dynamical evolution of a coalescing compact object binary system.

The effective one-body (EOB) theory based on the post-Newtonian (PN) approximation introduced by Buonanno and Damour~\cite{Damour1999} is a novel approach to studying the late dynamical evolution of a coalescing binary system. The EOB theory can provide an estimate of the gravitational waveform emitted throughout the inspiral, plunge, and coalescence phases for spinless and spin binaries \cite{Damour2000,Damour2000(2), Damour2001, Damour2006}. By calibrating the EOB model to numerical relativity simulations, EOB waveforms were improved, which are applied to the GW data analysis~\cite{Cook,Pan,Pan1,Damour2008,Damour20082, Boyle,Damour20083,Pan2,Pan3,Damour2007,Damour2009,Pan4,Bar,Pan5,Pan2014,Bohe,Damour2015,Cao2017}.

After the great success of the EOB theory based on PN approximation, in 2016, Damour~\cite{Damour2016} developed the EOB theory with post-Minkowskian (PM) approximation, in which the assumption that $v/c$ should be a small quantity was released. Since then, the correlational research based on the EOB theory with PM approximation has attracted great attention \cite{Damour2017, Damour2018,Damour2018new,Antonelli2019,Damour2019, Damour2020,HeLin2016, Blanchet2018, Cheung2018,Vines2019,Cristofoli2019, Collado2019, Bern2019,Bern20192, Plefka2019,BiniDamour2020,Cheung2020}.

The Hamilton equations \cite{Damour2000} of an EOB system based on the PN/PM approximation show that, for a self-consistent effective one-body (SCEOB) theory, all quantities appearing in equations should be constructed from the same physical model. Moreover, as we will show in this paper, the waveform should also be based on the same physical model.
That is, the Hamiltonian, radiation-reaction force (RRF), and waveform should be constructed in terms of the same effective spacetime.

To determine the expressions of the RRF and waveform for the ``plus" and ``cross" modes of GWs, we should first find the decoupled and variable separable equation for $\psi^B_{4}$ in the effective spacetime. Recently, we ~\cite{Jing} have successfully derived the decoupled equations of $\psi^{B}_{4}$ for even and odd parities in the Regge–Wheeler gauge~\cite{Thompson} by dividing the perturbation part of the metric into odd and even parities. The decoupled equations can be used to study the RRF and waveform in the effective spacetime and to set up an SCEOB theory. The explicit calculations for this model, however, are arduous tasks because we have to simultaneously solve two equations for the odd and even parities. Moreover, the equation for the odd parity is a third-order differential equation.

In this study, we derive another new decoupled and variable separable equation for $\psi^B_{4}$ in the effective spacetime and obtain the corresponding formal solution. To do so, we first take a gauge in which $\psi_{1}^{B}$ and $\psi_{3}^{B}$ vanish. This task can be done because, in a linear perturbation theory, $\psi_{0}^{B}$ and $\psi_{4}^{B}$ are gauge invariant, whereas $\psi_{1}^{B}$ and $\psi_{3}^{B}$ are not~\cite{Chandrasekhar}. In this gauge, the decoupled equation for $\psi^{B}_{4}$ can be obtained. Then, we separate variables for the decoupled equation of $\psi_{4}^{B}$ and obtain a formal solution. Based on this solution, we present the formulas for the RRF and waveform. Then, we set up an SCEOB theory for the spinless binaries, which is appropriate for any PM order. The theory not only releases the assumption that $v/c$ should be a small quantity but also resolves the contradiction that the Hamiltonian, RRF, and waveform are constructed from different physical models in the EOB theory with PN approximation. Compared with our previous SCEOB theory \cite{Jing}, the computational effort for the RRF and waveform in the new SCEOB theory was tremendously reduced.

The rest of the paper is organized as follows: In Sec. II, we derive a decoupled and variable separable equation for $\psi^B_{4}$ and determine the corresponding formal solution in the effective spacetime. In Sec. III, we set up the SCEOB theory for spinless binaries based on PM approximation. In Sec. IV, the final conclusions are presented.


\vspace{0.3cm}

\section{Formal solution for $\psi^B_{4}$ }

In an SCEOB theory, $\psi^B_{4}$ plays a central role in determining the RRF and waveform for the ``plus" and ``cross" modes of GWs. Therefore, we will first derive the decoupled and variables separable equation of motion and then present the corresponding formal solution for $\psi^B_{4}$ in the effective spacetime, which is described by \cite{Jing}:
\begin{eqnarray}
ds_{\rm eff}^2=g^{\text{eff}}_{\mu\nu}d x^\mu d x^\nu=\frac{\Delta_{r}}{r^2} dt^2-\frac{r^2}{\Delta_{r}}dr^2- r^2(d\theta^2+\sin^2\theta d\varphi^2),\label{Mmetric}
\end{eqnarray}
with
\begin{eqnarray}
&& \Delta_{r}=r^2- 2 GM_0 r+\sum_{i=2}^\infty a_i \frac{(GM_0\big)^i}{r^{i-2}},
\end{eqnarray}
where the definition for all parameters can be found in Ref. \cite{Jing}. In particular, the effective spacetime (\ref{Mmetric}) is type $D$ \cite{Carmeli}.

\subsection{Decoupled equation for $\psi^B_4$}

To decouple $\psi^B_{4}$ in the effective spacetime, we take the null tetrads for the spacetime (\ref{Mmetric}) as
\begin{eqnarray}
\nonumber &l^{\mu}&=\Big\{\frac{r^2}{\Delta_{r}}, 1 ,0,0\Big\}, \\
\nonumber &n^{\mu}&=\Big\{\frac{1}{2},-\frac{\Delta_{r}}{2 r^2},0,0\Big\}, \\
\nonumber &m^{\mu}&=\frac{1}{\sqrt{2}r }\Big\{0,0,1,\frac{{i}}{\sin\theta}\Big\}.\label{TeT}
\end{eqnarray}
Then, we have
\begin{eqnarray}\label{spin}
\rho=-\dfrac{1}{r}, \quad\mu=-\dfrac{\Delta_{r}}{2 r^3}, \quad\gamma=\dfrac{\Delta_{r}^{\prime}}{4r^2}+\mu, \quad\alpha=-\dfrac{\cot\theta}{2\sqrt{2} r } , \quad\beta=\dfrac{\cot\theta}{2\sqrt{2} r },
 \end{eqnarray}
here and hereafter, $^\prime$ represents a derivative with respect to $r$, and all other spin coefficients are equal to zero.

Ref. \cite{Carmeli} shows that there are three equations related to $\psi_{4}$:
\begin{align}
&\Delta\lambda-\overline{\delta}\nu=-(\mu+\overline{\mu})\lambda-(3\gamma-\overline{\gamma})\lambda+(3\alpha+\overline{\beta})\nu-\psi_{4}, \label{Eq1}\\
 \nonumber&\overline{\delta} \psi_{3}-D\psi_{4}+\overline{\delta} \phi_{21}-\Delta \phi_{20}
 =3 \lambda \psi_{2}-2 \alpha \psi_{3}-\rho \psi_{4}-2 \nu \phi_{10}+2 \lambda \phi_{11}\nonumber \\ & \ \ \ \ +(2 \gamma-2 \overline{\gamma}+\overline{\mu}) \phi_{20}-2\alpha\phi_{21}-\overline\sigma \phi_{22}, \label{Eq2} \\ &\Delta \psi_{3}-\delta \psi_{4}+\overline{\delta} \phi_{22}-\Delta \phi_{21} =3 \nu \psi_{2}-2(\gamma+2 \mu) \psi_{3}+4 \beta \psi_{4}-2 \nu \phi_{11}-\overline{\nu} \phi_{20}\nonumber \\ &\ \ \ \ +2 \lambda \phi_{12}+2(\gamma+\overline{\mu}) \phi_{21}-2( \overline{\beta}+ \alpha) \phi_{22}.\label{Eq3}\end{align}
We \cite{Jing} have shown that, in the effective background spacetime, the gravitational perturbation described by
\begin{align}\label{Pmetric}
 g_{\mu \nu}=g^{\text{eff}}_{\mu\nu}+\varepsilon h_{\mu \nu}^{B}\;,
\end{align}
can be achieved by perturbing all null tetrad quantities. Then, from Eqs. (\ref{Eq1}), (\ref{Eq2}) and (\ref{Eq3}), we derive the following perturbation equations:
\begin{align}\label{Eq11}
 &\psi_{4}^{B}+(\Delta+3\gamma-\overline{\gamma}+\mu+\overline{\mu})\lambda^{B}-(\overline{\delta}+3\alpha+\overline{\beta})\nu^{B}=0,\\
 &\nu^{B} (3\psi_{2}-2\phi_{11})-(\Delta+2\gamma+4\mu)\psi_{3}^{B}+(\delta+4\beta)\psi_{4}^{B}\nonumber \\&\ \ \ \ +(\Delta+2\gamma+2\overline{\mu})\phi_{21}^{B}-(\overline{\delta}+2\alpha+2 \overline{\beta})\phi_{22}^{B}=0,\label{Eq22}\\
 &\lambda^{B}(3\psi_{2}+2\phi_{11})-(\delta+2\alpha)\psi_{3}^{B}+(D-\rho)\psi_{4}^{B}\nonumber \\&\ \ \ \ -(\overline{\delta}+2\alpha)\phi_{21}^{B}+(\Delta+2\gamma -2\overline{\gamma}+\overline{\mu})\phi_{20}^{B}=0,\label{Eq33}\end{align}
where all quantities without and with the superscript $B$ represent the background and perturbation quantities, respectively.

Chandrasekhar pointed out that, for a linear perturbation given by (\ref{Pmetric}), $\psi_{0}^{B}$ and $\psi_{4}^{B}$ are gauge invariant, whereas $\psi_{1}^{B}$ and $\psi_{3}^{B}$ are not~\cite{Chandrasekhar}. Thus, we can choose a gauge in which $\psi_{1}^{B}$ and $\psi_{3}^{B}$ vanish without affecting $\psi_{0}^{B}$ and $\psi_{4}^{B}$. In this gauge, from Eqs. (\ref{Eq22}) and (\ref{Eq33}), we get $\nu^{B}$ and $\lambda^{B}$:
\begin{align}
&\nu^{B}=\frac{-(\Delta+2\gamma+2\overline{\mu})\phi_{21}^{B}-(\delta+4\beta)\psi_{4}^{B}+(\overline{\delta}+2\alpha+2\overline{\beta})\phi_{22}^{B}}{3\psi_{2}-2\phi_{11}},\label{lameq} \\
 &\lambda^{B}=\frac{(\overline{\delta}+2\alpha) \phi_{21}^{B}-(D-\rho)\psi_{4}^{B}-(\Delta+2\gamma-2\overline{\gamma}+\overline{\mu})\phi_{20}^{B}}{3\psi_{2}+2\phi_{11}}. \label{nulameq}
\end{align}
Substituting Eqs.~\eqref{lameq} and \eqref{nulameq} into Eq.~\eqref{Eq11}, we obtain the decoupled equation for $\psi_{4}^{B}$:
\begin{align}\label{decouple}
&\Big[(\Delta+3 \gamma-\overline{\gamma}+\mu+\overline{\mu}-F_{1})(D-\rho)
-F_{2}(\overline{\delta}+3 \alpha+\overline{\beta}-F_{3})(\delta+4 \beta)-F_4\Big]\psi_{4}^{B}=T_4^B,
\end{align}
with
\begin{align}\label{T4}
\nonumber T_4^B&=\Big((\Delta+3 \gamma+\mu-\overline{\gamma}+\overline{\mu})-F_{1}\Big)\Big[(\overline{\delta}+2\alpha-2\overline{\tau})\phi_{21}^{B}-(\Delta+2 \gamma-2\overline{\gamma}+\overline{\mu})\phi_{20}^{B}\Big]\\
 &-F_{2}\Big((\overline{\delta}+3 \alpha+\overline{\beta})-F_{3} \Big)\Big[(\overline{\delta}+2 \alpha+2 \overline{\beta})\phi_{22}^{B}-(\Delta+2 \gamma+2 \overline{\mu}) \phi_{21}^{B}\Big],
\end{align}
where the functions $F_i$ ($i=1,2,3,4$) are defined as
\begin{align}
 &F_{1}=-\dfrac{\Delta_{r}}{2r^2 (3\psi_{2}+2\phi_{11})} \dfrac{\partial}{\partial r}\big(3\psi_{2}+2\phi_{11}\big),\;\;\;\;\;\;F_{2}=\dfrac{3\psi_{2}+2\phi_{11}}{3\psi_{2}-2\phi_{11}},\nonumber\\ &F_{3}=\dfrac{1}{\sqrt{2} r (3\psi_{2}-2\phi_{11})}\dfrac{\partial}{\partial \theta}\big(3\psi_{2}-2\phi_{11}\big),\;\;\;\;\;\;\;\,F_{4}=3\psi_{2}+2\phi_{11}. \label{Ffuncs}
\end{align}

We aim to study a general perturbation as a superposition of waves with different frequencies ($\sigma^{\dagger}$) and periods ($2 m \pi$, $m=0,1,2,....$) in $\varphi$, so the perturbation in the $t$ and $\varphi$ directions may be written as $e^{i(\sigma^{\dagger} t + m \varphi)}$.

To solve Eq. (\ref{decouple}), we first introduce new derivative operators:
\begin{align}
\nonumber &\mathscr{D}_{n}=\partial_{r}+\frac{i r^2\sigma^{\dagger} }{\Delta_{r}}+n\frac{\Delta'_{r}}{\Delta_{r}}, \qquad\mathscr{D}_{n}^{\dagger}=\partial_{r}-\frac{i r^2\sigma^{\dagger}}{\Delta_{r}}+n\frac{\Delta'_{r}}{\Delta_{r}}, \\
 &\mathscr{L}_{n}=\partial_{\theta}+\frac{m}{\sin\theta}+n\cot\theta\qquad\mathscr{L}_{n}^{\dagger}=\partial_{\theta}-\frac{m}{\sin\theta}+n\cot\theta.
\end{align}
Then, the intrinsic derivatives, appearing in Eqs. (\ref{decouple}) and (\ref{T4}), may be rewritten as
\begin{align}\label{derivative}
D=\mathscr{D}_{0}, \qquad &\Delta=-\frac{\Delta_{r}}{2r^2 }\mathscr{D}_{0}^{\dagger}, \qquad\delta=\frac{1}{\sqrt{2} r}\mathscr{L}_{0}^{\dagger}, \qquad\overline\delta=\frac{1}{\sqrt{2}r}\mathscr{L}_{0}.
\end{align}
From Eqs.~\eqref{spin} and~\eqref{derivative} and by taking $\psi_{4}^{B}=r^{-4}\phi_{4}^{B}$, the decoupled GW equation~\eqref{decouple} becomes
\begin{align}\label{EqT}
\Big[\Delta_{r}\Big(\mathscr{D}_{-1}^{\dagger}+\frac{2r^2}{\Delta_{r}}F_1 \Big)\Big(\mathscr{D}_{0}-\frac{3}{r}\Big)+F_{2}\Big(\mathscr{L}_{-1}-\sqrt{2} r F_3\Big)\mathscr{L}_{2}^{\dagger}+2 r^2 F_{4}\Big]\phi_{4}^{B}=\mathscr{T}_4,
\end{align}
where
\begin{align}
\nonumber \mathscr{T}_4&=-2 r^6 T_4^B \\ \nonumber &= F_{2}r^4\Big(\mathscr{L}_{-1}-\sqrt{2} r F_{3} \Big)\mathscr{L}_{0}\phi_{22}^{B}+\frac{\Delta_{r}^{2} r^2 }{2}\Big(\mathscr{D}_{0}^{\dagger}+\frac{2}{r}+\frac{2 r^2}{\Delta_{r} }F_{1}\Big)\Big(\mathscr{D}_{0}^{\dagger}+\frac{1}{r}\Big)\phi_{20}^{B}\\ &+\frac{\Delta_{r}r^3}{\sqrt{2}}\Big[\Big(\mathscr{D}_{-1}^{\dagger}+\frac{3}{r}+\frac{2 r^2}{\Delta_{r}} F_{1}\Big) \mathscr{L}_{-1}+F_{2} \Big(\mathscr{L}_{-1}-\sqrt{2} r F_{3}\Big)\Big (\mathscr{D}_{-1}^{\dagger}+\frac{4}{r} \Big)\Big]\phi_{21}^{B}.\label{T4T}
\end{align}

Of note, although we decouple $\psi_{4}^{B}$ by taking the gauge in which $\psi_{1}^{B}$ and $\psi_{3}^{B}$ vanish, Eq. (\ref{EqT}) with $a_i=0$ ($i\ge 2$) is the same as that of the Schwarzschild spacetime obtained in Ref. \cite{Teukolsky}.

\vspace{0.3cm}

\subsection{Separation of the variables for $\psi^B_4$ }

For the spinless case, the non-vanishing trace-free Ricci and Weyl tensors in the null tetrads (\ref{TeT}) are
\begin{align}\label{pp}
 &\phi_{11}=\frac{1}{8 r^4 }\Big[4\Delta_{r}-4   r \Delta_{r}^{\prime} +  r^2 \Delta_{r}^{\prime\prime}+2 r^2\Big], \nonumber \\
 &\psi_{2}=\frac{1}{12 r^4 }\Big[12\Delta_{r}-6  r \Delta_{r}^{\prime} + r^2 \Delta_{r}^{\prime\prime}-2 r^2\Big],
 \end{align}
which indicate that, from Eq.~\eqref{Ffuncs}, $F_3=0$ and all other functions $F_j \, (j=1,2,4)$ are only $r$ dependent, so the variables of $\phi_{4}^{B}$ may be separated. To achieve this goal, we expand $\phi_{4}^{B}$ in terms of $_{-2} Y_{l m}(\theta )$:
\begin{equation}\label{phi4}
\phi_{4}^{B}=\sum_{ l m} \frac{1}{\sqrt{2\pi} } \int d\omega e^{-i(\omega t-m\varphi )}\,_{-2} Y_{l m}(\theta ) R _{l  m \omega }(r),
\end{equation}
where we have taken $\sigma^{\dagger} =-\omega$, and the functions $_{-2} Y_{l m}(\theta )$ may be normalized as
\begin{equation}\label{nor}
\int_{0}^{\pi}\,_{-2}Y_{\ell m}^{*}(\theta) \,_{-2}Y_{\ell m}(\theta) \sin\theta \mathrm{d}\theta =1.
\end{equation}
To compare our results with those obtained by Sasaki and Tagoshi \cite{Sasaki}, we take the normalized (\ref{nor}) instead of the usually normalized $\int \,_{-2}Y_{\ell m}^{u*}(\theta) \,_{-2}\,Y_{\ell m}^u(\theta) {d}\Omega =1 $, which shows that $ \,_{-2}Y_{\ell m}^{u}(\theta) =\,_{-2}\,Y_{\ell m}(\theta)/\sqrt{2\pi} $. Then, we obtain, from Eqs.~\eqref{EqT} and (\ref{T4T}), the separated radial equation:
\begin{eqnarray} \label{EqTS}
\left[r^{3}F_{4} \Delta_{r}^{2} \frac{\mathrm{d} }{\mathrm{d} r}  \Big(\frac{1}{r^{3} F_{4} \Delta_{r}} \frac{\mathrm{d} }{\mathrm{d} r}  \Big) + V (r) \right] R _{lm\omega}(r)= T _{\ell m \omega}(r),
\end{eqnarray}
with
\begin{align} \label{VrS}
&V (r)= \frac{r^{2} \omega(r^{2} \omega+2 i \Delta_{r}') }{\Delta_{r}} -i r \omega \Big(5  +\frac{2  r^{3}  F_1  }{\Delta_{r}} \Big)+\frac{3 (\Delta_{r}+ r \Delta_{r}')}{r^{2}} -6 r F_{1}+2 r^2 F_{4}-\lambda F_{2}, \\
&T _{l m \omega}(r)=\frac{1 }{2\pi }  \int_{-\infty }^{+\infty} dt \int d\Omega \ \mathscr{T}_{4} \ e^{i(\omega t-m \varphi)} \frac{ \ _{-2}Y^*_{l m}(\theta ) }{\sqrt{2\pi} }, \label{VrST}
\end{align}
where $\lambda $ is the eigenvalue of$\ _{-2}Y_{l m}(\theta )$, and
\begin{align}
\nonumber\mathscr{T}_{4}&=4\pi \Bigg\{ r^4 F_2 \ \mathscr{L}_{-1}\Big[\mathscr{L}_{0} \Big( T _{nn}\Big)\Big] +\frac{\Delta_{r}^{2}}{2} F_4 \ \mathscr{D}_{0}^{\dagger}\Big[\frac{r}{F_4}\mathscr{D}_{0}^{\dagger}\Big(
r T _{\overline{m}  \overline{m}}\Big)  \Big]\\   &+ \frac{\Delta_{r}^2}{\sqrt{2}}\  F_4\ \mathscr{D}_{0}^{\dagger}\Big[
\frac{r^3 }{\Delta_{r}F_4 }
\mathscr{L}_{-1}\Big(
T _{\overline{m} n}\Big) \Big]+\frac{\Delta_{r}^2}{\sqrt{2}} \ \frac{F_2 }{ r}
\mathscr{L}_{-1}\Big[
\mathscr{D}_{0}^{\dagger}\Big(
\frac{r^4 }{\Delta_{r}}T _{ \overline{m} n}\Big)  \Big]\Bigg\}.
\label{T4TTS}
\end{align}
In Eq.~\eqref{T4TTS}, we introduce $T_{n n}$, $T_{\overline{m} \overline{m}}$ and $T_{\overline{m} n}$, which are
\begin{equation}
T_{n n}=\dfrac{1}{4\pi}\phi^B_{22},\;\;\;\;\;\;T_{\overline{m}\overline{m}}=\dfrac{1}{4\pi}\phi^B_{20},\;\;\;\;\;\;T_{\overline{m}n}=\dfrac{1}{4\pi}\phi^B_{21}.\nonumber
\end{equation}

\subsection{Formal solution for $\psi_4^B$ in the effective spacetime}
In this subsection, we solve Eq. (\ref{EqTS}) using the Green function method. The homogeneous solutions for Eq.~\eqref{EqTS} can be expressed as
\begin{eqnarray}
R^{\rm in}_{\ell m\omega}
\to \left\{\begin{array}{cc}
B^{\rm trans}_{\ell m\omega}\Delta_r^2  e^{-i \omega r^*},
\ \ \ \ \ \ \ \ \ \ \ \ \ \ \ \ \ \ \ \ \ \ \ \ \  & \text{for} \ \ r\to r_+, \ \ \ \  \\
r^3 B^{\rm  ref}_{\ell m\omega}e^{i\omega r^*}+
r^{-1}B^{\rm inc}_{\ell m\omega} e^{-i\omega r^*},
\ \ \ \ \ \ & \text{for} \ \  r\to +\infty,
\end{array}\right.
\label{Kk}
\end{eqnarray}

\begin{eqnarray}
R^{\rm up}_{\ell m\omega}
\to \left\{\begin{array}{cc}
C^{\rm  up}_{\ell m\omega} e^{i \omega r^*}+
\Delta_r^2 C^{\rm ref}_{\ell m\omega} e^{-i\omega r^*},
\ \ \ \ \   & \text{for} \ \ r\to r_+, \ \ \ \  \\
C^{\rm trans}_{\ell m\omega} r^3  e^{i\omega r^*},
\ \ \ \ \ \ \ \ \ \ \ \ \ \ \ \ \ \ \ \ \ \ \  & \text{for} \ \  r\to +\infty,
\end{array}\right. ,
\label{Kkk}
\end{eqnarray}
where $r^*$ is the standard tortoise coordinate defined by
$
r^{*}=\int \frac{r^2}{ \Delta_r}dr
$.
Then, the inhomogeneous solution of the radial equation~\eqref{EqTS}, based on the causality property of waves generated by a source, may be constructed as
\begin{equation}
R_{\ell m\omega}=\frac{1}{2 i \omega C^{\rm trans}_{\ell m\omega}
     B^{\rm inc}_{\ell m\omega}}
 \left\{R^{\rm up}_{\ell m\omega}\int^r_{r_+}d\tilde{r}
\frac{ R^{\rm in}_{\ell m\omega}(\tilde{r} )
T_{\ell m\omega}(\tilde{r} ) }{\tilde{r} ^3 F_4(\tilde{r} )  \Delta_r^{2} (\tilde{r} ) }
+ R^{\rm in}_{\ell m\omega}\int^\infty_{r}d\tilde{r}
\frac{R^{\rm up}_{\ell m\omega}(\tilde{r} )  T_{\ell m\omega}(\tilde{r} ) }{\tilde{r} ^3 F_4(\tilde{r} )   \Delta_r^{2}(\tilde{r} ) }\right\}.
\end{equation}
Therefore, the solution for the radial equation (\ref{EqTS}) at the horizon is
\begin{equation}
R_{\ell m\omega}(r\to r_+)\to
\frac{{B^{\rm trans}_{\ell m\omega} \Delta_r^2 e^{-i \omega r^*}} }{
2i\omega C^{\rm trans}_{\ell m\omega}B^{\rm inc}_{\ell m\omega}}
\int^{\infty}_{r_+}d\tilde{r}  \frac{R^{\rm up}_{\ell m\omega}(\tilde{r} )
    T_{\ell m\omega}(\tilde{r} ) }{\tilde{r} ^3 F_4(\tilde{r} ) \Delta_r^{2}(\tilde{r} ) }
\equiv \tilde{Z}^{\rm H}_{\ell m\omega} \Delta_r^2 e^{-i \omega r^*}\,,
\label{Horizon}
\end{equation}
whereas the counterpart at the infinity is
\begin{equation}
R_{\ell m\omega}(r\to\infty)
 \to\frac{r^3e^{i\omega r^*} }{ 2i\omega
   B^{\rm inc}_{\ell m\omega}}
\int^{\infty}_{r_+}d\tilde{r}  \frac{T_{\ell m\omega}(\tilde{r} )
R^{\rm in}_{\ell m\omega}(\tilde{r} )
}{ \tilde{r} ^3 F_4(\tilde{r} )  \Delta_r^{2}(\tilde{r} )}
\equiv \tilde Z_{\ell m\omega} r^3 e^{i\omega r^*}\,.
\label{Infinfty}
\end{equation}

The energy momentum tensor for the EOB theory, i.e., a particle with the mass $m_0$ orbits around a massive black hole described by the effective metric, takes the form \cite{Sasaki81}
 \begin{equation}
T^{\mu\nu}
=\frac{m_0}{ r^2 \sin\theta dt/d\tau}\frac{d x^{\mu}}{ d\tau}
\frac{d x^{\nu}}{ d\tau}\delta(r-r(t))\delta(\theta-\theta(t))
  \delta(\varphi-\varphi(t)),\label{tijsl}
  \end{equation}
where $x^\mu=\bigl(t,\,r(t),\,\theta(t),\,\varphi(t)\bigr)$ is a geodesic
trajectory and $\tau=\tau(t)$ is the proper time along the geodesic.
By means of Eqs.~(\ref{T4TTS}), (\ref{tijsl}), and (\ref{VrST}) and performing the integration by parts, we obtain
\begin{eqnarray}
&&T _{l m \omega}(r)
=
-\frac{4m_0}{\sqrt{2\pi}}\int^{\infty}_{-\infty}
dt\int d\theta e^{i\omega t-im\varphi(t)} \nonumber\\
&&\ \ \ \
\times\Bigg\{-\frac{1}{ 2}\mathscr{L}_1^{\dag} \Big( \mathscr{L}_2^{\dag} \ _{-2}Y_{\ell m}(\theta)
\Big)
C _{n n} r^4 F_2\,\delta(r-r(t))
\delta(\theta-\theta(t)) \nonumber\\
&& \ \ \ \
+\frac{\Delta_{r}^2  }{ \sqrt{2} r }
\Big((1+F_2) \mathscr{L}_2^{\dag} \ _{-2}Y_{\ell m}(\theta) \Big)
\mathscr{D}_{0}^{\dagger} \Big[
\frac{C _{{\overline m} n} r^{4}}{\Delta_{r}}
\delta(r-r(t))\delta(\theta-\theta(t)) \Big] \nonumber\\
&& \ \ \ \
+\frac{1}{ 2\sqrt{2} }
\Big[ F_4  \frac{\partial }{\partial r}\Big(\frac{1}{r  F_4}\Big) \Big] \Big(\mathscr{L}_2^{\dag} \ _{-2}Y_{\ell m}(\theta)  \Big) C _{{\overline m} n}\Delta_{r} \,r^{4}\,
\delta(r-r(t))\delta(\theta-\theta(t)) \nonumber\\
&& \ \ \ \
-\frac{1}{ 4} \Delta_{r}^2 \ _{-2}Y_{\ell m}(\theta)  F_4 \mathscr{D}_{0}^{\dagger}\Big[\frac{r }{F_4}
\mathscr{D}_{0}^{\dagger} \Big(r C _{{\overline m}{\overline m}}
\delta(r-r(t))\delta(\theta-\theta(t))\Big) \Big]
\Bigg\}.
\label{Tsl}
\end{eqnarray}
with
\begin{eqnarray}
C _{n n}&=&\frac{\Delta_{r}}{ 4 r^4 E }\left(E
+ \frac{dr}{ d\tau} \right)^2,\nonumber\\
C _{{\overline m} n}&=&
\frac{\Delta_{r} }{ 2\sqrt{2} r^5 E}\left(E
+\frac{dr}{ d\tau} \right)
\left(\frac{ i L }{ \sin\theta}
-r^2 \frac{d\theta}{d\tau}\right),
\nonumber\\
C _{{\overline m} {\overline m}}&=&
\frac{\Delta_{r} }{ 2r^6 E }
\left(r^2 \frac{d\theta}{d\tau}-\frac{ i L }{ \sin\theta}
\right)^2.
\label{cijsl}
\end{eqnarray}
Eq.~(\ref{Tsl})  can be rewritten  as
\begin{eqnarray}
T _{\ell m\omega}(r)&=&-m_0 \int^{\infty}_{-\infty}dt
e^{i\omega t-i m \varphi(t)}
\Delta_{r}^2\Big\{  A_0 \delta(r-r(t)) +\Big[ A_1
\delta(r-r(t))\Big]^\prime
\nonumber\\
&+&\Big[ A_2
\delta(r-r(t))\Big]^{\prime\prime}\Big\},
\label{TgenTTsl}
\end{eqnarray}
with
\begin{eqnarray}
 A_0 &=&A _{nn\,0}+A _{{\overline m}n\,0}+
A _{{\overline m}{\overline m}\,0},\nonumber \\
 A_1 &=&A _{{\overline m}n\,1}+A _{{\overline m}{\overline m}\,1},\nonumber \\
 A_2 &=&A _{{\overline m}{\overline m}\,2},
\end{eqnarray}
where
\begin{eqnarray}
A _{nn\,0}&=&-\frac{2\,r^4 }{ \sqrt{2\pi}\,\Delta_{r}^2}\,
C _{n n}  \,F_2 \,
\mathscr{L}_1^+\Big[\mathscr{L}_2^+\Big(\ _{-2}Y_{\ell m}(\theta)\Big)  \Big],\nonumber \\
A _{{\overline m}n\,0}&=&\frac{r^3 }{ \sqrt{\pi}\Delta_{r}}
\,  C _{{\overline m} n} \Big[\big(1+F_2\big)\frac{ i r^2\omega }{ \Delta_{r}}+\frac{F_2}{r} -F_{2}^\prime-\frac{F_4^\prime}{F_4}\Big] \mathscr{L}_2^{\dag} \Big(\ _{-2}Y_{\ell m}(\theta) \Big),\nonumber \\
A _{{\overline m}{\overline m}\,0}
&=&\frac{r^2 }{ \sqrt{2\pi}}\,
C _{{\overline m} {\overline m}}\ _{-2}Y_{\ell m}(\theta)  \Bigl[
i\Bigl(\frac{ r^2\omega }{ \Delta_{r}}\Bigr)^\prime+\frac{r^4\omega^2 }{ \Delta_{r}^2}+ \frac{ i r^2\omega }{ \Delta_{r}}\Big(\frac{1}{r}+\frac{F_4^\prime}{F_4}\Big)+\frac{F_4^\prime}{r F_4}+\Big(\frac{F_4^\prime}{F_4}\Big)^\prime\Bigr],\nonumber \\
A _{{\overline m}n\,1}&=&\frac{
 r^3}{ \sqrt{\pi}\Delta_{r} }\,C _{{\overline m} n}
\big(1+F_2\big) \mathscr{L}_2^{\dag}\Big(\ _{-2}Y_{\ell m}(\theta) \Big)
,\nonumber \\
A _{{\overline m}{\overline m}\,1}
&=&\frac{ r^2 }{ \sqrt{2\pi}}
\,
C _{{\overline m}{\overline m}}\ _{-2}Y_{\ell m}(\theta)
\Bigl( -2 i\frac{ r^2\omega}{ \Delta_{r}}+\frac{1}{r}+\frac{F_4^\prime}{F_4}\Bigr),\nonumber \\
A _{{\overline m}{\overline m}\,2}
&=&-\frac{r^2}{ \sqrt{2\pi}}\,
C _{{\overline m} {\overline m}}\ _{-2}Y_{\ell m}(\theta),
\label{Aijs}
\end{eqnarray}
Inserting Eq.~(\ref{TgenTTsl}) into Eq.~(\ref{Infinfty}), we obtain
\begin{eqnarray}
\tilde Z _{\ell m\omega}=
\frac{m_0}{2i\omega B^{\rm inc}_{\ell m\omega}}
\int^{\infty}_{-\infty}dt e^{i\omega t-i m \varphi(t)}
\Bigl[ A_0
\frac{R^{\rm in}_{\ell m\omega}}{r^3 F_4}
- A_1
\Big(\frac{R^{\rm in}_{\ell m\omega}}{r^3 F_4}\Big)' + A_2 \Big(\frac{R^{\rm in}_{\ell m\omega}}{r^3 F_4}\Big)'' \Bigr]_{r=r(t), \theta=\theta(t)}.\nonumber \\
\label{ZZSch}
\end{eqnarray}
Because we focus on circular orbits, $r(t)$ in Eq. (\ref{ZZSch}) is not related to the time, so we can take $r(t)=r_0$. On the geodesic
trajectory, we also have $\theta(t)=\theta_0$ and $\varphi(t)=\Omega\, t$, where $\Omega$ is the angular velocity. Then, carrying out the integration for Eq. (\ref{ZZSch}), we find
\begin{eqnarray}
\tilde Z _{\ell m\omega}&=&
\frac{\pi m_0}{i\omega B^{\rm inc}_{\ell m\omega}}
\Bigl[ A_0
\frac{R^{\rm in}_{\ell m\omega}}{r^3 F_4}- A_1
\Big(\frac{R^{\rm in}_{\ell m\omega}}{r^3 F_4}\Big)' + A_2 \Big(\frac{R^{\rm in}_{\ell m\omega}}{r^3 F_4}\Big)'' \Bigr]_{r_0,\theta_0}\delta(\omega-\omega_n)\,,
\label{ZZZSch}
\end{eqnarray}
where $\omega_n=m\,\Omega$.
Then, noting the definition $\psi_4^B=\phi^B_4/r^4$ and using Eqs.~(\ref{phi4}), (\ref{Infinfty}) and (\ref{ZZZSch}), we find that the formal solution for $\psi_4^B$ is given by
\begin{eqnarray}
\psi^B_4&=&\frac{1}{ r}\sum_{\ell m n}
\frac{\pi m_0}{i\omega_n B^{\rm inc}_{\ell m\omega_n}}
\Bigl[ A_0
\frac{R^{\rm in}_{\ell m\omega_n}}{r^3 F_4}- A_1
\Big(\frac{R^{\rm in}_{\ell m\omega_n}}{r^3 F_4}\Big)' + A_2 \Big(\frac{R^{\rm in}_{\ell m\omega_n}}{r^3 F_4}\Big)'' \Bigr]_{r_0,\theta_0}\frac{{}_{-2}Y_{\ell m} }{ \sqrt{2\pi}}
e^{i\omega_n(r^*-t)+im\varphi}. \nonumber \\
\label{psi41}
\end{eqnarray}

\section{Framework of the self-consistent EOB theory for spinless binaries}
The basis of GWT is the late dynamical evolution of a coalescing compact object binary system. The dynamical evolution of the system is described by \cite{Damour2000,Jing}
\begin{eqnarray}
&&\frac{dr}{d t} - \frac{\pa H}{\pa P_r}
 = 0, \ \ \ \  \ \  \ \frac{d \varphi}{d t} - \frac{\pa H}{\pa P_\varphi}
 = 0, \nonumber \\ &&  \frac{d P_r}{d t} + \frac{\pa H}{\pa r}
= {\cal F}_r ,  \ \ \ \ \frac{d P_\varphi}{d t} = {\cal F}_\varphi. \label{HEqq}
\end{eqnarray}
Eq.~(\ref{HEqq}) shows that, for an SCEOB theory, $H$, $ {\cal F}_r $, and $ {\cal F}_\varphi $ and the waveform should be based on the same physical model.

With the effective line element (\ref{Mmetric}), energy map, and solution for $\psi_4^B$ (\ref{psi41}) at hand, we now set up an SCEOB theory for a spinless coalescing compact object binary system. That is,
we present the expressions of the Hamiltonian, RRF, and waveform for the system in the following.


 First, using Buonanno and Damour's proposals~\cite{Damour1999,Damour2000} and the energy map Eq. (7) in Ref. \cite{Jing}, we know that the Hamiltonian in Eq. (\ref{HEqq}) is still described by Eq. (10) in Ref. \cite{Jing}, which shows that the Hamiltonian is related to the effective spacetime described by Eq. (\ref{Mmetric}).


Second, the general relation between the flux of the GW energy emitted to infinity and the RRF is described by
$
\frac{d  E}{d t}= \dot{R}\,{\cal F}_{R} +
\dot{\varphi}\,{\cal F}_{\varphi}\,.
$
Buonanno and Damour \cite{Damour2000} showed that, for quasi-circular orbits, $ {\cal F}_{R}$ turns into zero, so we have
\begin{equation}
{\cal F}_{\varphi}= \frac{1}{\dot{\varphi}}\,\frac{d  E}{d t}.\label{FFF}
\end{equation}
Furthermore, the flux can be shown by~\cite{Ref:poisson,TagoshiSasaki745}:
\begin{eqnarray}
\frac{dE}{dt}=\frac{1}{16\pi G}\int (\dot{h}_{+}^2+\dot{h}_{\times}^2)r^2 d \Omega.\label{FG}
\end{eqnarray}
Noting $\psi^B_4=\frac{1}{2}(\ddot h_{+}-i\ddot h_{\times})$ at infinity and using Eq. \eqref{psi41}, we can get the gravitational waveform $h_{+}-ih_{\times}$ at infinity. By substituting $h_{+}-ih_{\times}$ into Eq. (\ref{FG}), we can obtain the expression for the flux of the GW energy emitted to infinity. Then, using Eq. (\ref{FFF}), we find that the RRF appearing in Eq. (\ref{HEqq}) is given by
\begin{eqnarray}
{\cal F}_{\varphi}= \frac{1}{\dot{\varphi}}\sum_{\scriptstyle \ell=2}^{\infty}\sum_{m=1}^\ell
\dfrac{2 \pi m_0^2 }{ G\omega_n^4}\Bigg| \frac{1}{B^{\rm inc}_{\ell m\omega_n}}
\Bigl[ A_0
\frac{R^{\rm in}_{\ell m\omega_n}}{r^3 F_4}- A_2
\Big(\frac{R^{\rm in}_{\ell m\omega_n}}{r^3 F_4}\Big)' + A_2 \Big(\frac{R^{\rm in}_{\ell m\omega_n}}{r^3 F_4}\Big)'' \Bigr]_{r_0,\theta_0}\Bigg|^2,
 \label{FFdE1}
\end{eqnarray}
which indicates that the RRF is constructed in terms of the effective spacetime.

Finally, by comparing $\psi^B_4$ with the waveform described by \cite{Kidder}
\begin{align}
h_{+}-i h_{\times}=\sum_{l=2}^{\infty} \sum_{m=-l}^{l} h^{lm}\frac{ \  _{-2}Y^{lm}(\theta, \varphi)}{\sqrt{2\pi}} ,\label{hh}
\end{align}
we find
\begin{eqnarray}
h^{l m}=\frac{1}{ r }
\frac{2 \pi m_0}{i\omega_n^3 B^{\rm inc}_{\ell m\omega_n}}
\Bigl[ A_0
\frac{R^{\rm in}_{\ell m\omega_n}}{r^3 F_4}- A_1
\Big(\frac{R^{\rm in}_{\ell m\omega_n}}{r^3 F_4}\Big)' + A_2 \Big(\frac{R^{\rm in}_{\ell m\omega_n}}{r^3 F_4}\Big)'' \Bigr]_{r_0,\theta_0} e^{i\omega_n(r^*-t)}.
\label{hform}
\end{eqnarray}
Clearly, the calculation for the waveform $h^{l m}$ is also based on the effective spacetime.

To improve the waveform, Damour and Nagar proposed [32] that, based on the waveforms obtained using the PN perturbation, multipolar waveforms should be built as
\begin{equation}\label{hlm}
h_{\ell m}=h_{\ell m}^{(N,\epsilon_p)}\,
\hat{S}_{\rm eff}^{(\epsilon_p)}\,T_{\ell m}\,e^{i\delta_{\ell m}}\,f_{\ell m}\,,
\end{equation}
where $h_{\ell m}^{(N,\epsilon_p)}$ is the leading Newtonian order term, $\hat{S}_{\rm  eff}^{(\epsilon_p)}$ is the relativistic conserved energy or angular momentum of the effective moving source, $T_{\ell m}$ resumes an infinite number of leading logarithms entering the tail effect, $e^{i\delta_{\ell m}}$ is an additional phase correction, and $f_{\ell m}$ is the remaining (essentially nonlinear) PN effects. Furthermore, to reproduce effects in the numerical simulations that go beyond the quasi-circular motion assumption, Pan et al \cite{Pan3} introduced a non-quasicircular (NQC) effect in $h_{lm}$.

We believe that the waveform Eq. (\ref{hform}) can be improved using Damour–Nagar–Pan's proposals \cite{Damour2007,Damour2009,Pan2,Pan3}.

\vspace{0.3cm}

\section{Conclusions and discussion}

The basis of GWT is the late dynamical evolution of a coalescing compact object binary system. To study the dynamical evolution, Buonanno et al. presented, based on the PN/PM approximation, a novel approach to map the two-body problem onto an EOB problem. Based on the Hamilton equations describing the dynamical evolution, in an SCEOB theory, the Hamiltonian $H$, RRF $ {\cal F}_\varphi $, and waveform should be constructed from the same physical model.

To build such an SCEOB theory, the key step is to find the decoupled equation with separable variables for $\psi^{B}_{4}$ in the effective background spacetime. Chandrasekhar pointed out~\cite{Chandrasekhar} that, for a linear gravitational perturbation described by Eq.~(\ref{Pmetric}), $\psi_{0}^{B}$ and $\psi_{4}^{B}$ are gauge invariant, whereas $\psi_{1}^{B}$ and $\psi_{3}^{B}$ are not. Therefore, we can take a gauge in which $\psi_{1}^{B}$ and $\psi_{3}^{B}$ vanish without affecting $\psi_{0}^{B}$ and $\psi_{4}^{B}$. In this gauge, we obtained the decoupled equation for $\psi^{B}_{4}$ in the effective spacetime. Then, we separated variables for the decoupled equation and obtained a formal solution of $\psi^B_4$. Based on the solution, we presented the formulas for the RRF and waveform.

With the Hamiltonian Eq. (10) in Ref. \cite{Jing}, RRF (\ref{FFdE1}) and waveform (\ref{hform}), which are based on the same effective spacetime, at hand, we set up an SCEOB theory for spinless binaries, which can be applicable to any PM orders. The theory not only releases the assumption that $v/c$ should be a small quantity but also resolves the contradiction that the Hamiltonian, RRF, and waveform are constructed from different physical models in the EOB theory based on PN approximation. Compared with our previous SCEOB theory \cite{Jing}, the computational effort for the RRF and waveform in the new SCEOB theory will be tremendously reduced.

Generally, the usefulness of an EOB model critically depends on the accuracy of the waveform template. Next, we will construct the GWT based on the SCEOB theory for the spinless binaries and compare it with that of the numerical relativity. We expect that the advantages of the SCEOB theory can improve the accuracy of the GWT. Furthermore, we will extend the study to spin binaries.

\vspace{0.3cm}

\vspace{0.3cm}
\acknowledgments
{ We would like to thank professors S. Chen and Q. Pan  for useful discussions on the manuscript. This work was supported by the Grant of NSFC Nos. 12035005, 12122504 and 11875025, and National Key Research and Development  Program of China No. 2020YFC2201400.}  

\end{document}